\documentclass[bimj,fleqn]{w-art}

\newcommand{\Var}{\operatorname{{\it Var}}}

\newcommand{\nfrac}{\nicefrac}

\newcommand{\whtau}{\widehat{\tau}}
\newcommand{\whtheta}{\widehat{\theta}}
\newcommand{\whq}{\widehat{q}}
\newcommand{\wttheta}{\widetilde{\theta}}

\usepackage{amsmath} 
\usepackage{times}
\usepackage{w-thm}
\usepackage{booktabs}
\usepackage{color}
\usepackage{longtable}
\usepackage{bbm}
\usepackage{amssymb}
\usepackage{multirow}
\usepackage{graphicx}
\usepackage[nice]{nicefrac}
\usepackage{lipsum}
\usepackage{enumitem}
\usepackage{array}
\usepackage{titlesec}
\usepackage[authoryear]{natbib}
\theoremstyle{plain}

\theoremstyle{definition}
\chardef\bslash=`\\ 

\newcommand{\citeintext}[1]{\citeauthor{#1} (\citeyear{#1})}
\newcommand{\citebrackets}[1]{(\citeauthor{#1}, \citeyear{#1})}

\begin{document}
\DOIsuffix{DOI}
\Volume{XX}
\Issue{ZZ}
\Year{2004}
\pagespan{1}{}
\keywords{Mann-Whitney effect $\theta$; Brunner-Munzel test; Studentized permutation test; Birnbaum-Klose inequality; $C^2$-test\\
\noindent \hspace*{-4pc} {\small\it}\\
\hspace*{-4pc} {\small\it }\\[1pc]
\noindent\hspace*{-4.2pc} Supporting Information for this article is available from the author or on the WWW under\break \hspace*{-4pc} \underline{http://dx.doi.org/10.1022/bimj.XXXXXXX} (please delete if not
applicable)
}  


\title[A New Approach to the nonparametric Behrens-Fisher Problem]{A New Approach to the Nonparametric Behrens-Fisher Problem with Compatible Confidence Intervals}

\author[Stephen Schüürhuis {\it{et al.}}]{Stephen Schüürhuis\footnote{Corresponding author: {\sf{e-mail: stephen.schueuerhuis@charite.de}}}\inst{,1}} 
\author[]{Frank Konietschke\inst{1}}
\author[]{Edgar Brunner\inst{2}}

\address[\inst{1}]{Charité - Universitätsmedizin Berlin, Corporate member of Freie Universität Berlin and Humboldt-Universität zu Berlin, Institute of Biometry and Clinical Epidemiology, Berlin, Germany}
\address[\inst{2}]{University of Göttingen, Department of Medical Statistics, Göttingen, Germany}
\Receiveddate{zzz} \Reviseddate{zzz} \Accepteddate{zzz}

\begin{abstract}
We propose a new test to address the nonparametric Behrens-Fisher problem involving different distribution functions in the two samples. Our procedure tests the null hypothesis $\mathcal{H}_0: \theta = \nicefrac{1}{2}$, where $\theta = P(X<Y) + \nicefrac{1}{2}P(X=Y)$ denotes the Mann-Whitney effect.  No restrictions on the underlying distributions of the data are imposed with the trivial exception of one-point distributions. The method is based on evaluating the ratio of the variance $\sigma_N^2$ of the Mann-Whitney effect estimator $\whtheta$ to its theoretical maximum, as derived from the Birnbaum-Klose inequality. Through simulations, we demonstrate that the proposed test effectively controls the type-I error rate under various conditions, including small sample sizes, unbalanced designs, and different data-generating mechanisms. Notably, it provides better control of the type-1 error rate compared to the widely used Brunner-Munzel test, particularly at small significance levels such as $\alpha \in \{0.01, 0.005\}$. Additionally, we derive range-preserving compatible confidence intervals, showing that they offer improved coverage over those compatible to the Brunner-Munzel test. Finally, we illustrate the application of our method in a clinical trial example.

\end{abstract}

\maketitle


\section{Introduction} 
\label{sec:introduction}
Comparing the means (expectations) of two independent groups with potentially different distributions (variances) is a fundamental problem in statistics, commonly encountered in fields such as medicine, biology, social sciences and education. When the assumption of equal variances between groups is questionable, the classical t-test may no longer be appropriate, leading to the well-known Behrens-Fisher problem. The classical Behrens-Fisher problem addresses this comparison under the assumption of unequal variances, typically using parametric methods that rely on normality. However, real-world data often violate these assumptions, with populations exhibiting skewed distributions or heavy tails, and these methods are also not suitable for ordinal data, such as 5-point Likert scales. Moreover, means fail to provide a meaningful definition of treatment effects. In those scenarios, it appears more appropriate to frame the problem as the \textit{nonparametric} Behrens-Fisher problem of testing
\begin{align*}
    \mathcal{H}_0: \theta = \text{P}(X_1 < X_2) + \nicefrac{1}{2}\text{P}(X_1 = X_2) = \nicefrac{1}{2},
\end{align*}
where $\theta$ represents the Mann-Whitney effect, a measure that offers a meaningful assessment of treatment effects, regardless of the underlying data distribution. \\

Many authors have developed methods for the nonparametric Behrens-Fisher problem, offering robust alternatives that relax the restrictive assumptions of parametric approaches (see e.g. \citeauthor{fligner1981}, \citeyear{fligner1981}; \citeauthor{brunner2000}, \citeyear{brunner2000}; \citeauthor{neubert2007}, \citeyear{neubert2007}; \citeauthor{pauly2016}, \citeyear{pauly2016}). These nonparametric methods are particularly valuable in settings like preclinical phases or rare diseases, where accurate comparisons between treatment groups with limited sample sizes are essential. Among various procedures, the test proposed by \citeintext{brunner2000} is widely recommended for comparing two heteroscedastic samples due to its accurate control of the type-I error rate for $\alpha \geq 0.05$ and sufficiently large sample sizes (\citeauthor{karch2021}, \citeyear{karch2021}; \citeauthor{wilcox2003applying}, \citeyear{wilcox2003applying}). However, the Brunner-Munzel test has two main weaknesses: I) it behaves liberally and over-rejects the null hypothesis at small sample sizes and/or low significance levels and II) its compatible confidence intervals are not range-preserving and show poor coverage for large effects $\theta$ (\citeauthor{noguchi2021}, \citeyear{noguchi2021}; \citeauthor{pauly2016}, \citeyear{pauly2016}). Those properties are particularly noteworthy given recent developments advocating for the adoption of more stringent significance levels (see, e.g., \citeauthor{johnson2013revised}, \citeyear{johnson2013revised}; \citeauthor{benjamin2018redefine}, \citeyear{benjamin2018redefine}; \citeauthor{held2019assessment}, \citeyear{held2019assessment}). It is the goal of this paper to address the limitations of the Brunner-Munzel test in detail, proposing a new closed-form procedure that improves performance also at small significance levels and provides range-preserving compatible confidence intervals. \\

Previous approaches addressing the nonparametric Behrens-Fisher problem include the tests by \citeintext{fligner1981} which is valid only for continuous distributions  and \citeintext{funatogawa2023}, who developed a testing procedure for ordered categorical data. In contrast, the Brunner-Munzel test imposes no restrictions on the underlying distributions and is often regarded as the preferred method (\citeauthor{karch2021}, \citeyear{karch2021}; \citeauthor{wilcox2003applying}, \citeyear{wilcox2003applying}), underpinned by being implemented in various software packages, such as \verb|rankFD| by \citeintext{konietschke2023}, \verb|lawstat| \citebrackets{lawstat} or \verb|brunnermunzel| \citebrackets{brunnermunzel}. In their paper, they propose approximating the distribution of the test statistic using a Satterthwaite-Welch-Smith $t$-approximation (see \citeauthor{satterthwaite1946}, \citeyear{satterthwaite1946}; \citeauthor{welch1937}, \citeyear{welch1937}; \citeauthor{smith1936}, \citeyear{smith1936}), which works well for sufficiently large sample sizes but may prove inadequate for small sample sizes and low significance levels. In response, \citeintext{neubert2007} proposed approximating the distribution of the test statistic using a studentized permutation approach, demonstrating improved type-I error rate control for small sample sizes and low significance levels (see \citeauthor{pauly2016}, \citeyear{pauly2016}; \citeauthor{noguchi2021}, \citeyear{noguchi2021}). However, permutation tests rely on resampling for p-value computation, which can be computationally intensive and limit their practicality in certain applications. Additionally, the associated confidence intervals are not necessarily range-preserving, requiring range-preserving transformations as proposed by \citeintext{pauly2016}. Yet, it seems reasonable to conclude that the Brunner-Munzel test, as a closed-form procedure, and the studentized permutation test, as a resampling method, currently serve as benchmark approaches for addressing the nonparametric Behrens-Fisher problem. \\

Conceptually, the Brunner-Munzel test and the corresponding studentized permutation test are similar, as both seek to correct for liberal tendencies by approximating the distribution of the test statistic: the former uses a $t$-approximation, while the latter employs a studentized permutation distribution. For both tests, the need for approximations partly arises from using the \citeintext{delong1988} variance estimator, which is known to be positively biased \citebrackets{brunner2024_2}. Moreover, the $t$-approximation of the Brunner-Munzel test statistic is theoretically arguable, as the numerator, which includes the asymptotically normal effect estimator, and the denominator, which contains the DeLong variance estimator, are not stochastically independent. Additionally, the distribution of the variance estimator generally only approximates a $\chi^2$-distribution, which can lead to poor performance, especially in settings with small sample sizes. \\

Here, we approach the nonparametric Behrens-Fisher problem from a different perspective: rather than proposing a new approximation of the distribution of the test statistic, we revisit the variance estimator and explore how it may be adjusted to counteract liberal behavior. More concretely, we propose a new test by considering the ratio of the variance of the Mann-Whitney effect estimator, $\sigma_N^2$, to its theoretical upper bound as derived from the Birnbaum-Klose inequality \citebrackets{birnbaum1957}, eventually resulting in an adjusted version of the Brunner-Munzel test statistic. Unlike existing approaches relying on the variance estimator by \citeintext{delong1988}, we employ the unbiased rank-based variance estimator introduced by \citeintext{brunner2024_2}. Additionally, we provide compatible confidence intervals for the new method and show that they are range-preserving. To narrow the focus of our study, we concentrate on tests with closed-form solutions, excluding resampling-based methods in the present paper. Using simulation, we illustrate that the new test procedure has a decent type-I error rate control while maintaining power comparable to the Brunner-Munzel and studentized permutation tests. Furthermore, we compare the coverage probability of the associated confidence intervals, illustrating that the new confidence intervals provide satisfactory coverage for an increased range of true effects $\theta$ as compared to the Brunner-Munzel test. \\

This paper is organized as follows: Section~\ref{sec:example} presents a motivating data example. The statistical model, estimators, and methods under consideration are detailed in Sections~\ref{sec:notation} and ~\ref{sec:tests}. A comprehensive simulation study targeting the type-I error rate and power of the tests, as well as the coverage probability of the compatible confidence intervals, is presented in Section~\ref{sec:simulation} The data example is revisited in Section~\ref{sec:application}. Finally, the paper closes with a discussion and an outlook on potential developments.

\section{Data Example}
\label{sec:example}
To illustrate the application of all the methods under consideration, we reanalyze the data from a shoulder tip pain study conducted by \citeintext{jorgensen1995} and later reported by \citeintext{lumley1996}, which was also used in the papers by \citeintext{brunner2000} and \citeintext{neubert2007}. This randomized controlled trial introduces a specific suction procedure to remove air from the abdomen as a new intervention. A total of n = 22 patients are randomized to the treatment group, while n = 19 patients are in the control group. Pain is assessed using a score ranging from 1 (no pain) to 5 (severe pain) on the evening of the second day post-surgery. That is, this trial's primary endpoint has an ordinal measurement level. For details on the procedure and the trial, we refer to \citeintext{lumley1996} and the original trial by \citeintext{jorgensen1995}. The following Table~\ref{tab:example1} summarizes the trial data utilizing absolute frequencies of the different pain score values. The boxplot illustrates the empirical distribution of the data. Note that the pain score is highly skewed for the intervention group.
\begin{table}[h!]
    \centering
    \begin{minipage}[t]{0.45\textwidth}
        \centering
        \vspace{23pt}
        \caption{Pain Scores on the Evening of Day 2 After Laparoscopic Surgery: Comparison Between 22 Patients with Specific Suction Procedure and 19 Control Patients}
        \begin{tabular}{lcccccc}
           \toprule
           & \multicolumn{5}{c}{Pain Score} & \\
           \cmidrule(lr){2-6}
           Treatment Arm & 1 & 2 & 3 & 4 & 5 & Total \\
           \midrule
           Specific Suction & 16 & 5 & 0 & 1 & 0 & 22 \\
           Control & 4 & 1 & 5 & 7 & 2 & 19 \\
           \bottomrule
        \end{tabular}
        \label{tab:example1}
    \end{minipage}
    \hfill
    \begin{minipage}[t]{0.45\textwidth}
        \centering
        \vspace{0pt}
        \includegraphics[scale = 0.355]{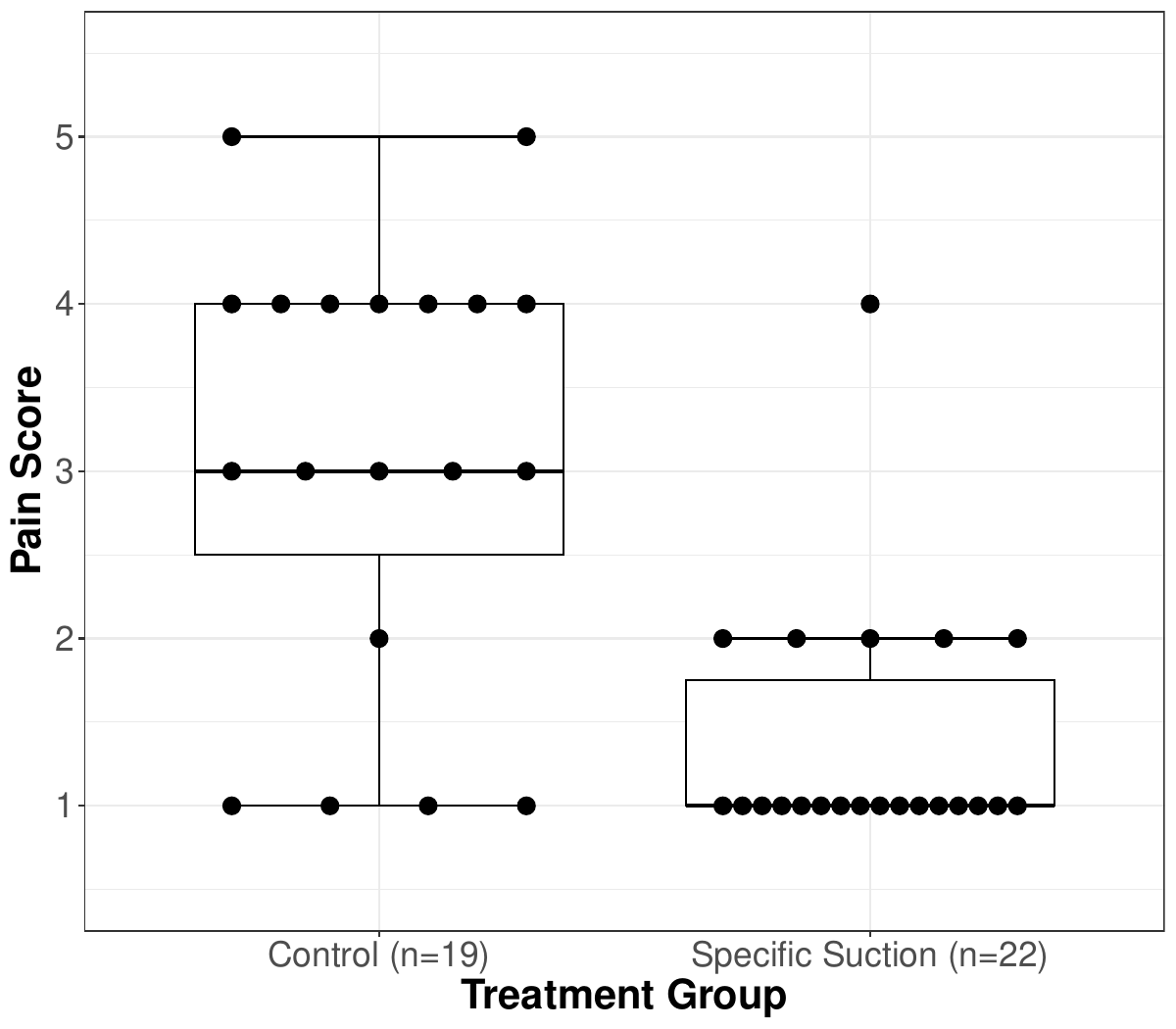}
        \label{fig:example1}
    \end{minipage}
\end{table}

\section{Notation and Nonparametric Model} \label{sec:notation}
\subsection{The Mann-Whitney Effect}
We consider the nonparametric model stated by \citeintext{brunner2000} and consider independent random variables  $X_{ik} \sim F_i, \quad k = 1, \dots, n_i$, where $i = 1,2$ denotes the treatment arm and $N = n_1 + n_2$ the total sample size. Note that the distributions $F_i$ may be arbitrary, excluding the case of one-point distributions. The estimand most naturally associated with the nonparametric Behrens-Fisher problem is the Mann-Whitney effect
\begin{align*}
    \theta = \int F_1 \, dF_2 = \text{P}(X_{11} < X_{21}) + \nicefrac{1}{2} \text{P}(X_{11} = X_{21}),
\end{align*}  
where $F_i = (F_i^+ + F_i^-)/2$ represents the normalized version of the cumulative distribution function \citebrackets{Ruy1980}. Note that this definition provides a unified notation that covers continuous and discrete data. In practice, $\theta$ can be interpreted as the probability that a random observation drawn from $F_2$ will be larger than a random observation drawn from $F_1$ while accounting for the probability of ties. That is, if $\theta > \, (<) \, 0.5$, $X_{11}$ is said to tend to smaller (larger) values than $X_{21}$. If $\theta=0.5$, $X_{11}$ and $X_{21}$ are said to be stochastically comparable. Correspondingly, the nonparametric Behrens-Fisher problem can be formulated in terms of the null hypothesis $\mathcal{H}_0: \theta = \nfrac{1}{2}$, representing the two-sided hypothesis of no tendency towards larger values in either population. A commonly used estimator $\whtheta_N$ of $\theta$ can be derived by replacing the theoretical distribution functions $F_i$ by their empirical counterparts $\widehat{F}_i$, resulting in 
\begin{align}
    \whtheta_N = \int \widehat{F}_1d\widehat{F}_2 = \frac{1}{n_1}\left(\bar{R}_{2\cdot} - \frac{n_2 + 1}{2}\right), \label{eq:theta}
\end{align}
where $\bar{R}_{2\cdot} = \frac{1}{n_2}\sum_{i=1}^{n_2}R_{ik}$ is the mean of all ranks $R_{ik}$ and $R_{ik}$ is the rank of observation $X_{ik}$ among all $i=1,\dots,N$ observations, see \citeintext{brunnerBook2019} (Section 3.3, Result 3.1) for details. For asymptotic results, we assume that $N/n_i$ is uniformly bounded, that is, $N/n_i \leq n_0 < \infty$, compare \citeintext{brunnerBook2019} for details (Section 3, Assumption 3.17 and 3.19). \\

The construction of tests for $\mathcal{H}_0: \theta = \nicefrac{1}{2}$ requires knowledge about the underlying variance $\sigma_N^2 ~{=}~\Var(\whtheta_N)$ or $v_N^2 {=}~\Var(\sqrt{N}\whtheta_N) = N\sigma_N^2$ of the estimator $\whtheta_N$. Many papers address the estimation of $\sigma_N^2$. For example, \citeintext{fligner1981} proposed an estimator valid only for continuous distribution functions. In contrast, \citeintext{bamber1975} provided an unbiased estimator that remains valid in the presence of ties but did not consider a rank-based representation of the estimator (for details, we refer to \citeintext{brunner2024_2}). The following section introduces the variance estimators used for the tests in this paper. Specifically, we present the estimator proposed by \citeintext{delong1988}, which is used in the Brunner-Munzel test, as well as the rank-based representation of the \citeintext{bamber1975} estimator provided by \citeintext{brunner2024_2}, which is used in our proposed test.

\subsection{Variance Estimation}
\subsubsection{Estimator by \citeintext{delong1988}} \label{sec:delong}
For the derivation of this variance estimator, \citeintext{brunner2000} used the asymptotic variance
\begin{align*}
    v_N^2 = N\left( \frac{\sigma_1^2}{n_1} + \frac{\sigma_2^2}{n_2} \right)
\end{align*}
based on the asymptotic equivalence theorem, as discussed in \citeintext{brunnerBook2019} (Section 7, Theorem 7.16) or \citeintext{brunner2000}. Here, $\sigma_1^2 = \Var(F_2(X_{1j}))$ and $\sigma_2^2 = \Var(F_1(X_{2i}))$ are generally unknown and need to be estimated from the data. With $R_{ik}^{(i)}$ denoting the internal rank of observation $X_{ik}$ within sample $i$ and 
\begin{align}
    R_{ik}^* = R_{ik} - R_{ik}^{(i)}
\label{eq:placement}
\end{align}
the so-called placement of $X_{ik}$ in the respective other sample $i' \neq i$ \citebrackets{orban1982}, replacing $F_i$ with the corresponding empirical distribution function $\widehat{F}_i$ leads to the $L_2$-consistent estimator
\begin{align*}
    \widehat{\sigma}_i^2 &= \frac{1}{n_i - 1} \sum_{k=1}^{n_i} \left( R_{ik}^* - \bar{R}_{i\cdot} + \frac{n_i + 1}{2} \right)
\end{align*}
of $\sigma_i^2$. The corresponding estimator for $v_N^2$ can eventually be obtained as the plug-in estimator
\begin{align}
    \widehat{v}_{DL}^2 &= N\left( \frac{\widehat{\sigma}_1^2}{n_1} + \frac{\widehat{\sigma}_2^2}{n_2} \right).
    \label{eq:delong}
\end{align}
Since the estimator provided by \citeintext{brunner2000} is identical to that of \citeintext{delong1988}, we refer to it as the DeLong estimator for the remainder of the paper. Note that this estimator may be biased for various underlying distributions, as discussed by \citeintext{brunner2024_2}.

\subsubsection{Estimator by \citeintext{brunner2024_2}}
In a recent paper, \citeintext{brunner2024_2} introduced a rank-based version of the unbiased variance estimator proposed by \citeintext{bamber1975}. By considering the $U$-statistic representation of $\whtheta_N$ instead of the rank-based form in \eqref{eq:theta}, they show that the true variance $\sigma_N^2$ can be expressed as
\begin{align*}
    \sigma_N^2 = \frac{1}{n_1n_2}\left\{(n_1-1)\sigma_2^2 + (n_2-1)\sigma_1^2 + \theta(1-\theta) - \frac{\tau}{4}\right\},
\end{align*}
where again $\sigma_1^2 = \Var(F_2(X_{1j}))$ and $\sigma_2^2 = \Var(F_1(X_{2i}))$. Additionally, the probability of ties in the overlap region of $F_1$ and $F_2$, denoted as $\tau := P(X_{1j} = X_{2i})$, can be estimated by  
\begin{align}
    \whtau_N = \frac{1}{n_1}\left\{\bar{R}_{2\cdot}^+ - \bar{R}_{2\cdot}^- - \left(\bar{R}_{2\cdot}^{(2)+} - \bar{R}_{2\cdot}^{(2)-} \right) \right\}
    \label{eq:tau}
\end{align}
where, $\bar{R}_{2\cdot}^+$ and $\bar{R}_{2\cdot}^+$ represent the means of the maximum and minimum ranks $R_{2k}^+$ and $R_{2k}^-$ among all $N$ observations and $\bar{R}_{2\cdot}^{(2)+}$ $\bar{R}_{2\cdot}^{(2)-}$ denote the means of the maximum and minimum internal ranks $R_{2k}^{(2)+}$ and $R_{2k}^{(2)-}$ among all $n_2$ observations in the second sample, respectively. For further details on different ranking approaches, we refer to \citeintext{brunnerBook2019} (Section 2, Definition 2.19). \\

Finally, the rank-based estimator $\widehat{\sigma}_N^2$ is given by 
\begin{align}
    \widehat{\sigma}_N^2 = \frac{1}{d_n} \left\{ \sum_{i=1}^2 \sum_{k=1}^{n_i} (R_{ik}^* - \bar{R}_{i\cdot}^*)^2 - n_1n_2\left( \whtheta_N(1-\whtheta_N) - \frac{\whtau_N}{4} \right) \right\},
    \label{eq:var_unb}
\end{align}
where $ d_n = n_1(n_1-1)n_2(n_2-1) $ and $ \bar{R}_{i\cdot}^* = \frac{1}{n_1} \sum_{i=1}^{n_i} R_{ik}^* $ represents the mean of the placements $ R_{ik}^*$ defined in \eqref{eq:placement} \citebrackets{brunner2024_2}.  \\

In contrast to the DeLong estimator $ \widehat{v}_{DL}^2 $, this estimator is unbiased and non-negative for sample sizes $ n_i \geq 2 $. Moreover, \citeintext{brunner2024_2} have shown that $ \widehat{\sigma}_N^2 $ is bounded by $ \whtheta(1 - \whtheta)/(m - 1) $, where $ m = \min\{n_1, n_2\} $. This sharp upper bound can be considered an empirical version of the Birnbaum-Klose inequality (\citeyear{birnbaum1957}). Given these preferable properties, we will use this variance estimator in our new test procedure.

\section{Tests for the Nonparametric Behrens-Fisher Problem} \label{sec:tests}
\subsection{The Brunner-Munzel-Test (2000)}
To derive tests for the nonparametric Behrens-Fisher Problem, first consider that under $\mathcal{H}_0: \theta = \nicefrac{1}{2}$, the statistic $\sqrt{N}(\whtheta_N - \nicefrac{1}{2})/v_N$ is asymptotically standard normal. For large sample sizes, it therefore follows that
\begin{align}
    \text{P}\left(|\sqrt{N}(\whtheta_N - \nicefrac{1}{2})/\widehat{v}_N| >  z_{1-\alpha/2}\right) \approx \alpha,
    \label{eq:C2-init}
\end{align}
where $\widehat{v}_N$ denotes some consistent estimator for $v_N = \sqrt{v_N^2}$. Now let $\widehat{v}_{DL}^2$ denote the DeLong estimator in Equation~\eqref{eq:delong}, then the test statistic
\begin{align}
      T_{N}^{BM} &= \sqrt{N}(\whtheta_N - \nicefrac{1}{2})/\widehat{v}_{DL} \quad  \xrightarrow{\quad d \quad} \quad   \mathcal{N}(0,1)
      \label{eq:BM-Stat}
\end{align}
becomes precisely the test statistic known from the Brunner-Munzel test. Simulation studies have demonstrated that large sample sizes are required for a satisfactory approximation. In many medical and biological applications, however, the number of available observations may be quite small. Consequently, \citeintext{brunner2000} proposed to approximate the distribution of $T_{N}^{BM}$ by a $t_{f}$-distribution, where the degrees of freedom are estimated through the Satterthwaite-Smith-Welch approximation
\begin{align*}
    \widehat{f} = \frac{\left\{\sum_{i=1}^2\widehat{\sigma}_i^2/(N-n_i)\right\}^2}{\sum_{i=1}^2\{\widehat{\sigma}_i^2/(N-n_i)\}^2/(n_i-1)}.
\end{align*}
Finally, the hypothesis $\mathcal{H}_0: \theta = \nfrac12$ is rejected at a two-sided significance level $\alpha$ if 
\begin{align}
   |T_N^{BM}| > t_{\widehat{f}, 1-\alpha/2} 
   \label{eq:BM-Test}
\end{align}
where $t_{\widehat{f}, 1-\alpha/2}$ denotes the $1-\alpha/2$-quantile of a central $t$-distribution with $\widehat{f}$ degrees of freedom. \\

\begin{figure}[h]
    \centering
    \includegraphics[scale = 0.4]{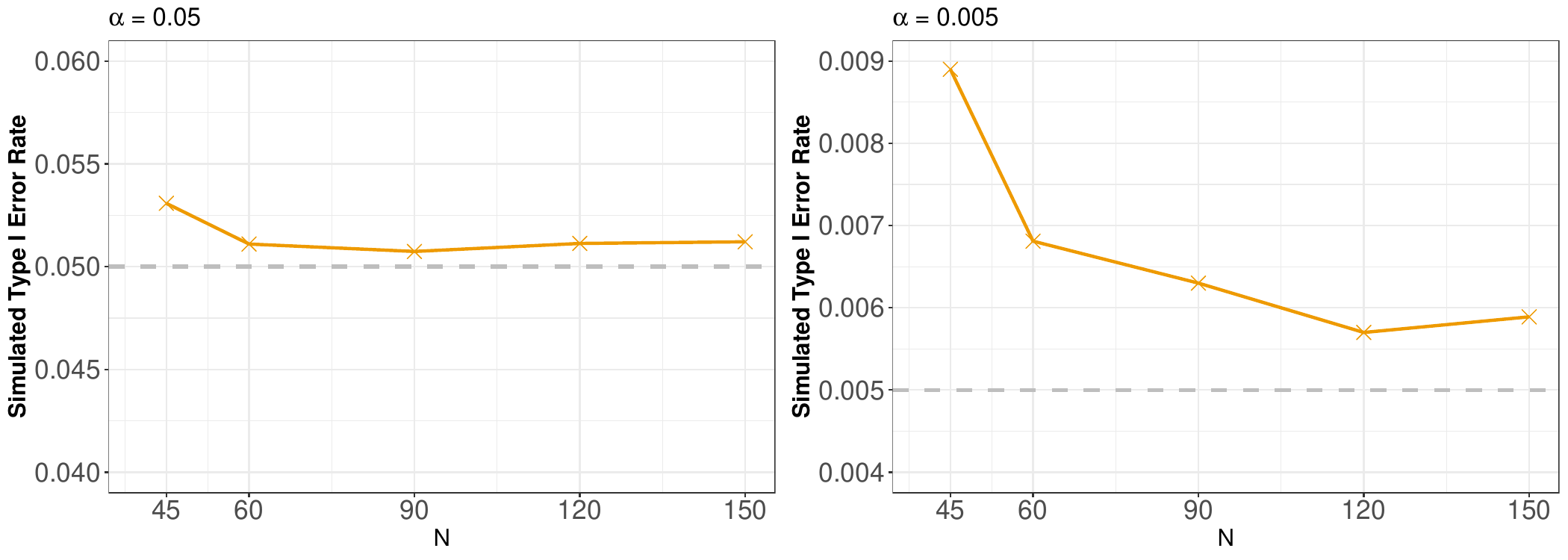}
    \caption{Simulated type-I error rate of the Brunner-Munzel test $T_N^{BM}$ for normal distributions $F_1 = \mathcal{N}(0, 1)$ and $F_2 = \mathcal{N}(0, 9)$ based on 100,000 replications at a two-sided nominal significance level for $\alpha = 0.05$ (left) and $\alpha = 0.005$ (right). The sample sizes are chosen such that $n_1 = 2n_2$, resulting in negatively paired variances. The dashed grey line represents the nominal significance level.}
    \label{fig:bm_demo}
\end{figure}
Exemplary results of the type-I error rate of the test are presented in Figure~\ref{fig:bm_demo}, considering the case of heteroscedastic normal samples $F_1 = \mathcal{N}(0,1)$ and $F_2 = \mathcal{N}(0,9)$ with negatively paired variances ($n_1 = 2n_2$). The results demonstrate that the Brunner-Munzel test may become slightly liberal for small sample sizes at a significance level of $\alpha = 0.05$. Notably, the test exhibits a clear liberal tendency at a lower $\alpha$-level of 0.5\%, consistent with the findings of \citeintext{noguchi2021}. This behavior suggests that the $t$-approximation may be inadequate in certain scenarios and motivates the exploration of alternative methods. \\

The $100(1-\alpha)\%$-confidence interval most naturally associated with this test can be obtained by inverting the test in \eqref{eq:BM-Test} to the confidence interval
\begin{align}
    \theta_{L, U}^{BM} = \whtheta_N \mp t_{\widehat{f},1-\alpha/2}\frac{\widehat{v}_{DL}}{\sqrt{N}}.
    \label{eq:BM-CI}
\end{align}
Although this confidence interval is, by construction, compatible to the test decision from the test in \eqref{eq:BM-Test}, it suffers from a poor coverage probability for large effects $\theta$ \citebrackets{pauly2016}. Additionally, the confidence interval is not necessarily range-preserving. To illustrate this drawback, consider the artificial dataset in Table~\ref{tab:example_bm}.
\begin{table}[h!]
    \centering
    \caption{Example dataset to illustrate the use of the presented confidence intervals}
    \begin{tabular}{lp{8.5cm}}
        \toprule
        \textbf{Groups} & \textbf{Outcome} \\
        \midrule
        Group 1 & 1956, 3828, 2051, 3721, 3233, 2000, 4000, 4428, 2603, 2370 \\
        \midrule
        Group 2 & 820, 3364, 1957, 1851, 2984, 744, 2044 \\
        \bottomrule 
    \end{tabular}
    \label{tab:example_bm}
\end{table}
Using this data, we obtain a Mann-Whitney effect estimate of $\whtheta = 0.8$ and a p-value of 0.0239, indicating the rejection of $\mathcal{H}_0: \theta = \nfrac{1}{2}$ at a two-sided significance level of $\alpha = 0.05$. The corresponding compatible 95\%-confidence interval calculated based on \eqref{eq:BM-CI} is $[0.55, 1.05]$, demonstrating a non-range preserving confidence interval ($1.05 > 1$). Alternatively, a range-preserving transformation-based confidence interval (see, e.g., \citeauthor{pauly2016}, \citeyear{pauly2016}) could be used. Those, however, are not compatible and may contradict the decision to reject the null hypothesis drawn from the Brunner-Munzel test. 

\subsection{The \citeintext{neubert2007}-Test}
To address the liberal behavior of the Brunner-Munzel test for small sample sizes and $\alpha \in \{0.005, 0.01\}$, \citeintext{neubert2007} proposed a studentized permutation test, see also \citeintext{pauly2016}.  The first step in conducting the test is to calculate the test statistic $T_{N}^{BM}$ from Equation~\eqref{eq:BM-Stat}. Afterwards, their approach involves collecting all data into a vector
\begin{align*}
    \mathbf{X} = (X_{11},\dots,X_{1n_1},X_{21},\dots,X_{2n_2}),
\end{align*}
randomly permute the data $h = 1,\ldots,n_{p}$ times and compute the Brunner-Munzel test statistic $T_{N,h}^{BM}$ for each permuted dataset. All test statistics are stored in a vector $\mathbf{T_{perm}^{BM}} = (T_{N,1}^{BM}, \ldots, T_{N,n_p}^{BM})$ and the two-sided p-value is computed as
    \begin{align*}
        p = 2\cdot min\left\{ \frac{1}{n_p}\sum_{h=1}^{n_p}\mathbbm{1}_{\{T_{N,h}^{BM}<T_{N}^{BM}\}} ,\frac{1}{n_p}\sum_{h=1}^{n_p}\mathbbm{1}_{\{T_{N,h}^{BM}>T_{N}^{BM}\}}\right\}.
    \end{align*}
Finally, the null hypothesis is rejected if the p-value obtained from the empirical permutation distribution of $\mathbf{T_{perm}^{BM}}$ is smaller than the nominal significance level of $\alpha$.  Based on the approximation of the test statistic using the permutation distribution, a $100(1-\alpha)\%$-confidence interval can be constructed as 
\begin{align}
    \theta_{L, U}^{p} = \left[\whtheta_N - c_1\frac{\widehat{\nu}_{DL}}{\sqrt{N}}, \whtheta_N - c_2\frac{\widehat{\nu}_{DL}}{\sqrt{N}}\right],
    \label{eq:Perm-CI}
\end{align}
where $c_1 = q_{p, 1-\alpha/2}$ and $c_2 = q_{p, \alpha/2}$ denote the empirical $1 - \alpha/2$ and $\alpha/2$-quantile of $\mathbf{T_{perm}^{BM}}$. \\

\citeintext{pauly2016} provided a rigorous proof stating that the conditional permutation distribution of the studentized test statistics is approximately standard normal for any value of $\theta$, justifying the construction of compatible confidence intervals. By simulation, \citeintext{neubert2007} already demonstrated that this test adequately controls the type-I error rate especially in small sample sizes, indicating that distribution of the permuted test statistics approximates the null distribution of $T_{N}^{BM}$ quite well. \citeintext{noguchi2021} illustrated that the studentized permutation test robustly controls the type-I error rate also for a nominal significance level as small as $\alpha = 0.005$. Therefore, we include this test in this paper as a well-known alternative to the standard Brunner-Munzel test, referring to it as the \citeintext{neubert2007}-Test in the following.\\

Since the test relies on iteratively permuting the data, it tends to be computationally intensive. While this is not an issue when analyzing a single dataset, simulation-based power evaluations and trial designs may involve relatively long runtimes. In this context, analytical solutions would offer an advantage. Moreover, the corresponding confidence interval tends to be liberal for large effects, such as $\theta > 0.8$, as noted by \citeintext{pauly2016}. In the next section, we will discuss a modified version of the Brunner-Munzel test which shall reduce the liberal behavior of the traditional test. To this end, we will use the unbiased variance estimator $\widehat{\sigma}_N^2$ from Equation~\eqref{eq:var_unb}, in conjunction with the \citeintext{birnbaum1957} inequality. The test will then be compared to the Brunner-Munzel test and the corresponding studentized permutation test by means of a simulation study.

\subsection{A New Test for $\mathcal{H}_0: \theta = \nicefrac{1}{2}$} \label{subsec:c2}
To derive the proposed test of the hypothesis $\mathcal{H}_0: \theta = \nicefrac{1}{2}$, we again start by noting that
\begin{align}
    \text{P}\left(|\whtheta_N-\theta|/\sigma_N >  z_{1-\alpha/2}\right) \approx \alpha
\end{align}
since $\whtheta_N$ is asymptotically normally distributed. Various options to approximate the unknown variance $\sigma_N^2$ in \eqref{eq:C2-init} are available. The most straightforward solution directly replaces $\sigma_N^2$ with a suitable variance estimator $\widehat{\sigma}_N^2$. For instance, replacing $\sigma_N^2$ with the estimator in \eqref{eq:delong}, one obtains the Brunner-Munzel test as presented earlier. Alternatively, one could approximate the unknown variance $\sigma_N^2$ using the theoretical upper bound of $\sigma_N^2$ which is given by $\sigma_{\text{max}}^2:= \theta(1-\theta)/m$ according to the Birnbaum-Klose inequality \citebrackets{birnbaum1957}. Here, $m = min\{n_1,n_2\}$ denotes the sample size of the smaller of both groups. While this approach does not require estimating the variance, the resulting test will, unsurprisingly, be rather conservative. \\

As a compromise between relying on an estimator $\widehat{\sigma}_N^2$ versus using a theoretical upper bound $\sigma_{\text{max}}^2$, one could consider some distance between the true variance $\sigma_N^2$ and $\sigma_{\text{max}}^2$ for deriving a hypothesis test. The option we consider here is using the ratio between the true variance $\sigma_N^2$ and the maximum variance $\sigma_{\text{max}}^2$.  Correspondingly, we define the ratio
\begin{align}
    r := \frac{\sigma_{\text{max}}^2}{\sigma_N^2} = \frac{\theta(1-\theta)}{m\sigma_N^2}.
    \label{eq:ratio}
\end{align}
Both the maximum variance $\sigma_{\text{max}}^2$ and the unknown true variance $\sigma_N^2$ can be estimated by substituting the estimators $\widehat{\theta}_N$ and $\widehat{\sigma}_N^2$ from Equations~\eqref{eq:theta} and \eqref{eq:var_unb}.
Hence, we can approximate the ratio in \eqref{eq:ratio} by
\begin{align}
    \widehat{r} = \frac{\whtheta_N(1-\whtheta_N)}{m\widehat{\sigma}_N^2}.
    \label{eq:ratio_est}
\end{align}
Assuming that $\widehat{r} \approx r$, the unknown variance $\sigma_N^2$ can be approximated by
\begin{align}
    &\frac{\whtheta_N(1-\whtheta_N)}{m\widehat{\sigma}_N^2} \approx \frac{\theta(1-\theta)}{m\sigma_N^2} \quad \Longrightarrow \quad \sigma_N^2 \approx  \frac{\theta(1-\theta)}{\whtheta_N(1-\whtheta_N)}\widehat{\sigma}_N^2
    \label{eq:approx}
\end{align} 
 \vspace{0.3cm}
\begin{remark}
In Equation~\eqref{eq:ratio_est}, an alternative approximation is given by 
$\widehat{r} = \frac{\whtheta_N(1-\whtheta_N)}{(m-1)\widehat{\sigma}_N^2}$
since ${\widehat{\sigma}_N^2 \leq \frac{\whtheta_N(1-\whtheta_N)}{(m-1)}}$ which may be regarded as the ``empirical Birnbaum-Klose'' inequality. Again replacing $\widehat{r} \approx r$ leads to the approximation $\sigma_N^2 \approx \frac{m-1}{m}\frac{\theta(1-\theta)}{\whtheta_N(1-\whtheta_N)}\widehat{\sigma}_N^2$, thereby inducing a dependency on the smaller of both groups. Simulations suggest that the resulting test does not demonstrate significant improvements. For the sake of brevity, we have chosen to omit this test for the remainder of the paper.
\end{remark} \vspace{0.3cm}

Plugging in approximation \eqref{eq:approx} into Equation~\eqref{eq:C2-init}, we find that
\begin{align}
    P\left(|\whtheta_N-\theta|/\sigma_N >  z_{1-\alpha/2}\right)  &= P\left(\frac{(\whtheta_N-\theta)^2}{\sigma_N^2} >  c_{1-\alpha}\right) \\ 
    &\approx P\left( \frac{(\whtheta_N-\theta)^2}{\widehat{\sigma}_N^2}\frac{\whtheta_N(1-\whtheta_N)}{\theta(1-\theta)} > c_{1-\alpha}\right) \approx \alpha,
    \label{eq:der_C2}
\end{align}
where $c_{1-\alpha} = z_{1-\alpha/2}^2$ denotes the $1-\alpha$-quantile of the $\chi_1^2$-distribution. Therefore, setting $\theta = \nicefrac{1}{2}$, an asymptotic test for $\mathcal{H}_0: \theta = \nicefrac{1}{2}$ can be obtained as
\begin{align}
    C^2 = 4 \frac{\whtheta_N(1-\whtheta_N)}{\widehat{\sigma}_N^2}(\whtheta_N-\nicefrac{1}{2})^2 \,\, \stackrel{\mathcal{H}_0}{\sim} \,\, \chi_1^2,
    \label{eq:c2_theta0.5}
\end{align}
which essentially is a corrected squared Brunner-Munzel test statistic, however, incorporating an unbiased variance estimator instead of the variance estimator by \citeintext{delong1988}. The hypothesis $\mathcal{H}_0: \theta=\nfrac{1}{2}$ is rejected at significance level $\alpha$ if $C^2 > c_{1-\alpha}$. The corresponding p-value is computed as $p = 1 - F_{\chi^2_1}(C^2)$,
where $F_{\chi^2_1}$ denotes the cumulative distribution function of a central $\chi^2_1$-distribution. For the remainder of the paper, we will refer to this test as $C^2$-test. \\

It remains to derive compatible confidence intervals for this test. From \eqref{eq:der_C2}, it directly follows that
\begin{align*}
    &\text{P}\left( \frac{(\whtheta_N-\theta)^2}{\widehat{\sigma}_N^2}\frac{\whtheta_N(1-\whtheta_N)}{\theta(1-\theta)} \leq c_{1-\alpha}\right) \approx 1 - \alpha,
\end{align*}
Following the ideas by \citeintext{wilson1927} in case of binomial proportions, an approximate $100(1-\alpha)\%$ confidence interval for $\theta$ can be obtained by solving the quadratic inequality
\begin{align*}
    &\quad \frac{(\whtheta_N-\theta)^2}{\widehat{\sigma}_N^2}\frac{\whtheta_N(1-\whtheta_N)}{\theta(1-\theta)} \leq c_{1-\alpha} \quad \Longleftrightarrow \quad
    (\whtheta_N-\theta)^2\leq \widehat{\sigma}_N^2\frac{c_{1-\alpha}}{\whtheta_N(1-\whtheta_N)} \theta(1-\theta)
\end{align*}
for $\theta$, leading to the confidence bounds 
\begin{align}
        \theta_{L,U}^{r}
        &= \frac{1}{2\left(1 + \whq c_{1-\alpha}\right)} \left(2\whtheta_N + \whq c_{1-\alpha} \mp \sqrt{\whq^2 c_{1-\alpha}^2 + 4\widehat{\sigma}_N^2 c_{1-\alpha}}\right),
        \label{eq:C2-CI}
\end{align}
where $\whq = \widehat{\sigma}_N^2/(\whtheta_N(1-\whtheta_N))$. Here, the index $r$ refers to the fact that we consider the ratio in Equation~\eqref{eq:ratio}. By construction, the confidence interval $\theta_{L,U}^r$ aligns with the decision derived from the test \eqref{eq:c2_theta0.5}, such that if $\theta \notin [\theta_L^r, \theta_U^r]$, it follows that $C^2 > c_{1-\alpha}$ and vice versa. For details on the derivation, we refer to Appendix A.1. \\

\begin{remark}
The confidence interval $\theta_{L,U}^r$ in Equation~\eqref{eq:C2-CI} is range-preserving by construction. The corresponding proof can be found in Appendix A.2. To illustrate this property, we revisit the data example in Table~\ref{tab:example_bm}. Applying the $C^2$-test to analyze the data, we obtain a p-value of 0.0282 and the $95$\% confidence interval $[0.53, 0.93]$, which is both range-preserving and compatible with the test decision.
\end{remark} \vspace{0.3cm}

In case of completely separate samples, we have $\whtheta_N \in \{0,1\}$ and $\widehat{\sigma}_N^2 = 0$, resulting in undefined values for the $C^2$-test statistic. However, this scenario may occur in simulations with small sample sizes even for moderate values of $\theta$. Therefore, it is necessary to address this situation separately. To this end, we propose using $\sigma_{\text{max}}^2 = \theta(1-\theta)/m $ instead of the ratio approximation. As a result, variance estimation is no longer required, leading to the simplified test statistic
\begin{align}
 C^2_{\sigma_{max}} = 4m (\whtheta_N-\nicefrac{1}{2})^2
 \label{eq:c2_theta=0}
\end{align}
By plugging in $\whtheta_N \in \{0,1\}$, this statistic becomes $C^2_{\sigma_{max}}  = m$ thus depending on ${m = min\{n_1,n_2\}}$ only. The p-value of this test would then be given by $p = 1 - F_{\chi^2_1}(m)$. Hence for, a significance level of $\alpha = 0.05$ or $\alpha = 0.01$, this test will consistently reject the null hypothesis $\mathcal{H}_0: \theta = \nicefrac{1}{2}$ when $m > 3$ or $m > 6$, respectively. This result seems reasonable, as the criterion for rejecting the null hypothesis is expected to be comparatively liberal when observing effects as extreme as $\widehat{\theta} \in \{0,1\}$. To obtain compatible confidence intervals, we again approximate the true variance through the upper bound $\sigma_{\text{max}}^2$ and solve
\begin{align*}
    (\whtheta_N-\theta)^2 \leq c_{1-\alpha}\theta(1-\theta)/m,
\end{align*}
for $\theta$, leading to the confidence interval
\begin{align}
    \theta_{L,U}^{BK} = \frac{1}{2(m+c_{1-\alpha})} \left(2m\whtheta_N + c_{1-\alpha} \mp \sqrt{4m\whtheta_N(1-\whtheta_N)c_{1-\alpha} + c_{1-\alpha}^2}\right) \label{eq:CI-BK}
\end{align}
analogously to the confidence interval in \eqref{eq:C2-CI}. Here, the index $BK$ refers to the fact that we use the bound of the Birnbaum-Klose inequality to replace the unknown variance \citebrackets{birnbaum1957}. With $\whtheta_N \in \{0,1\}$, those boundaries further simplify to
\begin{align}
\theta_{L}^{BK}(1) = \frac{m}{m + c_{1-\alpha}} \quad \text{and} \quad \theta_{U}^{BK}(1) = 1 \label{eq:CI_theta=1}
\end{align}
and
\begin{align}
\theta_{L}^{BK}(0) = 0 \quad \text{and} \quad \theta_{U}^{BK}(0) = \frac{c_{1-\alpha}}{m + c_{1-\alpha}}, \label{eq:CI_theta=0}
\end{align}
hence turning into one-sided confidence intervals for extreme values of $\whtheta$.  
\label{sec:methods}
\newpage

\section{Simulation Study}
\label{sec:simulation}
\citeintext{neubert2007} conducted extensive simulations demonstrating that the studentized permutation test effectively controls the preassigned type-I error level $\alpha$ under the null hypothesis, even for very small sample sizes, outperforming the Brunner-Munzel test in this regard. Additionally, they provided some power simulation results. In this section, we compare both approaches to the $C^2$-test by examining the type-I error rate and assessing the power of all three methods. Furthermore, we evaluate the coverage of the corresponding confidence intervals.

\subsection{Type-I Error Rate Simulation}
The distributions used for generating data considered in the type-I error simulation study are summarized in Table~\ref{tab:settings-t1e}. These settings encompass metric (discrete and continuous) and ordinal data. Additionally, homoscedastic and heteroscedastic variances, along with symmetric and skewed distributions, are considered. The Laplace distribution is included to also investigate heavy-tailed distributions.
\begin{table}[ht]
    \centering
    \caption{Distributions considered in the type-I error rate simulation study; All scenarios are chosen such that the null hypothesis that $\mathcal{H}_0: \theta = \nfrac{1}{2}$ holds.}
    \begin{tabular}{p{1cm}p{4cm}p{7cm}} \toprule
        \textbf{Setting} & \textbf{Distribution} & \textbf{Parameters} \\
        \midrule
1 & Normal distribution & $F_1 = \mathcal{N}(0,1)$ and $F_2 = \mathcal{N}(0,1)$ \\
2 & Normal distribution & $F_1 = \mathcal{N}(0,1)$ and $F_2 = \mathcal{N}(0,9)$ \\
\midrule
3 & Beta distribution & $F_1 = \mathcal{B}(1,1)$ and $F_2 = \mathcal{B}(1,1)$ \\
4 & Beta distribution & $F_1 = \mathcal{B}(1,1)$ and $F_2 = \mathcal{B}(5,5)$ \\
5 & Beta distribution & $F_1 = \mathcal{B}(2,5)$ and $F_2 = \mathcal{B}(2,5)$ \\
6 & Beta distribution & $F_1 = \mathcal{B}(5,5)$ and $F_2 = \mathcal{B}(1,1)$ \\
\midrule
7 &  Ordered categorical data & Based on $F_1 = \mathcal{B}(1,1)$ and $F_2 = \mathcal{B}(1,1)$ \\
8 & Ordered categorical data & Based on $F_1 = \mathcal{B}(1,1)$ and $F_2 = \mathcal{B}(5,5)$ \\
9 & Ordered categorical data & Based on $F_1 = \mathcal{B}(2,5)$ and $F_2 = \mathcal{B}(2,5)$ \\
10 & Ordered categorical data & Based on $F_1 = \mathcal{B}(5,5)$ and $F_2 = \mathcal{B}(2,2)$ \\
\midrule
11 & Poisson distribution & $F_1 = \mathcal{P}ois(1)$ and $F_2 = \mathcal{P}ois(1)$ \\
\midrule
12 & Exponential distribution & $F_1 = \mathcal{E}xp(1)$ and $F_2 = \mathcal{E}xp(1)$ \\
\midrule
13 & Laplace distribution & $F_1 = \mathcal{L}(0,1)$ and $F_2 = \mathcal{L}(0,1)$ \\
14 & Laplace distribution & $F_1 = \mathcal{L}(0,1)$ and $F_2 = \mathcal{L}(0,3)$ \\
\midrule
\bottomrule
\label{tab:settings-t1e}
    \end{tabular}
\end{table}
For details on simulating ordered categorical data, we refer to Appendix A.3. For the sample sizes, we consider the following balanced as well as unbalanced designs:
\begin{align*}
    (n_1, n_2) &\in \{(15,15), (30,30), (45,45), (60,60), (75,75)\} \\
    (n_1, n_2) &\in \{(15,30), (20,40), (30,60), (40,80), (50,100)\} \\
    (n_1, n_2) &\in \{(30,15), (40,20), (60,30), (80,40), (100,50)\}.
\end{align*}
That is, we consider the sample size ratios $n_1/n_2 \in \{0.5, 1, 2\}$, covering both positive pairing (larger variance in the larger group) and negative pairing (larger variance in the smaller group) for normal, beta, and laplace distributions. The nominal $\alpha$-level is set to $\alpha \in \{0.001,0.005,0.01,0.05\}$.  This approach is motivated by recent recommendations advocating for the use of more stringent significance levels, such as $0.5\%$ (see e.g. \citeauthor{johnson2013revised}, \citeyear{johnson2013revised}; \citeauthor{benjamin2018redefine}, \citeyear{benjamin2018redefine}; \citeauthor{held2019assessment}, \citeyear{held2019assessment}). Note that, to accommodate the asymptotic nature of the proposed method, we opted not to go below $n_i < 15$ in our simulation study. We use $n_{iter} = 100,000$ for all tests, giving rise to a Monte Carlo error of about $0.0001$ to $0.0007$ for $\alpha = 0.001$ to $\alpha = 0.05$. Finally, we estimate the type-I error rate as $\frac{1}{n_{iter}}\sum_{i=1}^{n_{iter}}\mathbbm{1}_{\{\phi_g(\mathcal{X}_i) = 1\}}$ where $\phi_g: \mathcal{X} \xrightarrow{\quad} \{0,1\}$ represents one of the statistical tests $g = \{C^2, T_N^{BM}, T_{perm}^{BM}\}$ considered throughout this manuscript. For the studentized permutation test, we set $n_{p} = 10,000$ as the number of permutations. \\

For all analyses, we used our implementations in the software package R (Version 4.1.2) in conjunction with the RStudio IDE (Version 2022.07.1) \citebrackets{R}. In the case of separate samples, the variance estimator $\widehat{\nu}_{DL}^2$ used for the Brunner-Munzel and the corresponding permutation test becomes zero such that corresponding tests and confidence intervals are undefined. Details on handling this singularity can be found in Appendix A.4. All code necessary to reproduce the presented results will be included in the supplementary material. \\

\begin{figure}[htbp]
    \centering
    \includegraphics[scale = 0.35]{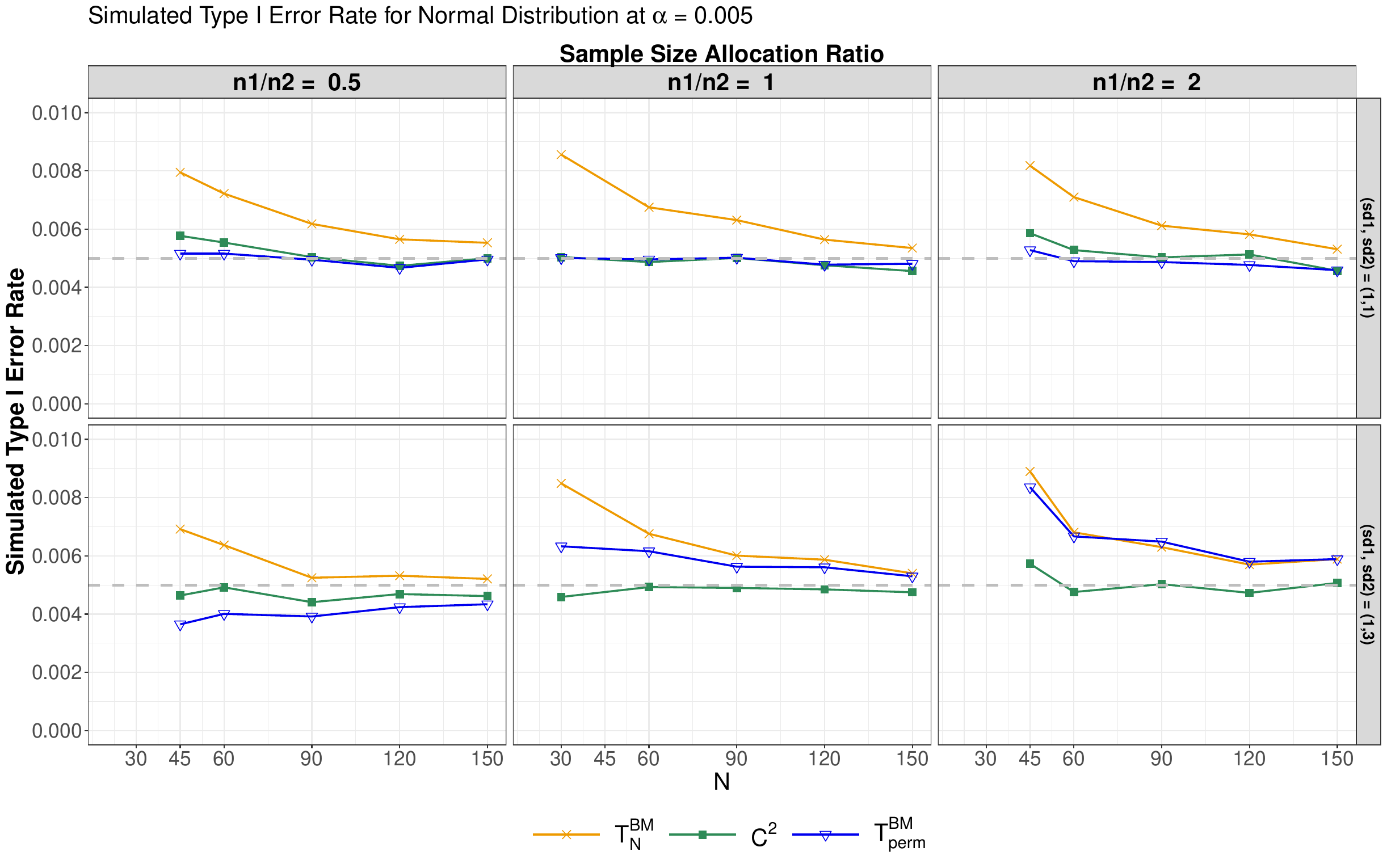}
    \caption{Type-I error rate for normal distributions $F_1 = \mathcal{N}(0, 1)$ and $F_2 = \mathcal{N}(0, \sigma_2^2)$ based on 100,000 replications at a two-sided nominal significance level of $\alpha = 0.005$ for all considered tests. The dashed grey line represents the nominal significance level.}
    \label{fig:normal_005}
\end{figure}

Exemplary results from the type-I error rate simulations for settings 1, 2, and 7–10 in Table~\ref{tab:settings-t1e} are illustrated in Figures~\ref{fig:normal_005} to \ref{fig:ordinal_005} for $\alpha = 0.005$. Additional simulation results are provided in the supplementary materials. Here, we briefly summarize and discuss the findings for all tests included in the simulation study, beginning with a description of the plot setup.
The columns of the plot represent different sample size settings. In the left column, $n_2 = 2n_1$; in the center column, $n_1 = n_2$; and in the right column, $n_1 = 2n_2$. The rows correspond to different parameter settings for the respective distributions. For example, in the lower row of Figure~\ref{fig:normal_005}, $F_1 = \mathcal{N}(0,1)$ and $F_2 = \mathcal{N}(0,9)$. Here, the lower right plot illustrates the case of negatively paired variances, while the lower left plot represents positive pairing. In all figures, the $x$-axis shows the total sample size $N = n_1 + n_2$, and the $y$-axis displays the simulated type-I error rate. The grey dashed line indicates the target nominal $\alpha$ level. Based on the presented figures and the comprehensive simulation study detailed in the supplementary material, the simulation results can be summarized as follows:
\begin{itemize}
\item The \textbf{Brunner-Munzel test} ($T_N^{BM}$), represented by the orange lines with crosses, demonstrates liberal behavior for small sample sizes across all examined scenarios, regardless of the underlying distribution. This tendency becomes more pronounced as the significance level decreases. However, as the sample size increases, the type-I error rate converges to the nominal $\alpha$-level. These findings align with previous research, such as \citeintext{noguchi2021} and \citeintext{pauly2016}, which similarly observed that the test exhibits a liberal tendency, particularly at smaller significance levels.
    \item The \textbf{studentized permutation test} ($T_{perm}^{BM}$), represented by the blue line with triangles, exhibits improved type-I error rate control across nearly all underlying distributions, particularly for small sample sizes. This suggests that the studentized permutation distribution effectively approximates the null distribution of the Brunner-Munzel test statistic, aligning with the theoretical foundation established by \citeintext{pauly2016}, who also demonstrated that the permutation test can robustly control the type-I error rate even for smaller sample sizes. However, it is important to note that when variances are either negatively or positively paired, the studentized permutation test appears to inherit the behavior of the original Brunner-Munzel test, exhibiting liberal or conservative tendencies, respectively. This pattern is evident, for instance, in the lower row of Figure~\ref{fig:normal_005}, as well as in cases involving heteroscedastic Beta distributions and corresponding 5-point Likert scales.
    \item The \textbf{$C^2$-test} (green line with squares) demonstrates improved type-I error rate control compared to the Brunner-Munzel test. This improvement generally holds across different underlying distributions and significance levels $\alpha$, suggesting that the $\chi_1^2$-approximation used in Equation~\eqref{eq:c2_theta0.5} is valid. The only exception arises in small sample scenarios involving highly skewed 5-point Likert-scale data, as illustrated in Figure~\ref{fig:ordinal_005}, where the underlying distribution is $F_i = \mathcal{B}(2,5), i = 1,2$. Here, the test slightly exceeds the target $\alpha$-level. This behavior may be attributed to the extreme skewness of the simulated data, where the discretization causes most values to fall into a single category. As a result, the data may be close to an empirical one-point distribution, causing variance estimation to degenerate and leading to liberal type-I error rates.
\end{itemize}

\begin{figure}[htbp]
    \centering
    \includegraphics[scale = 0.4]{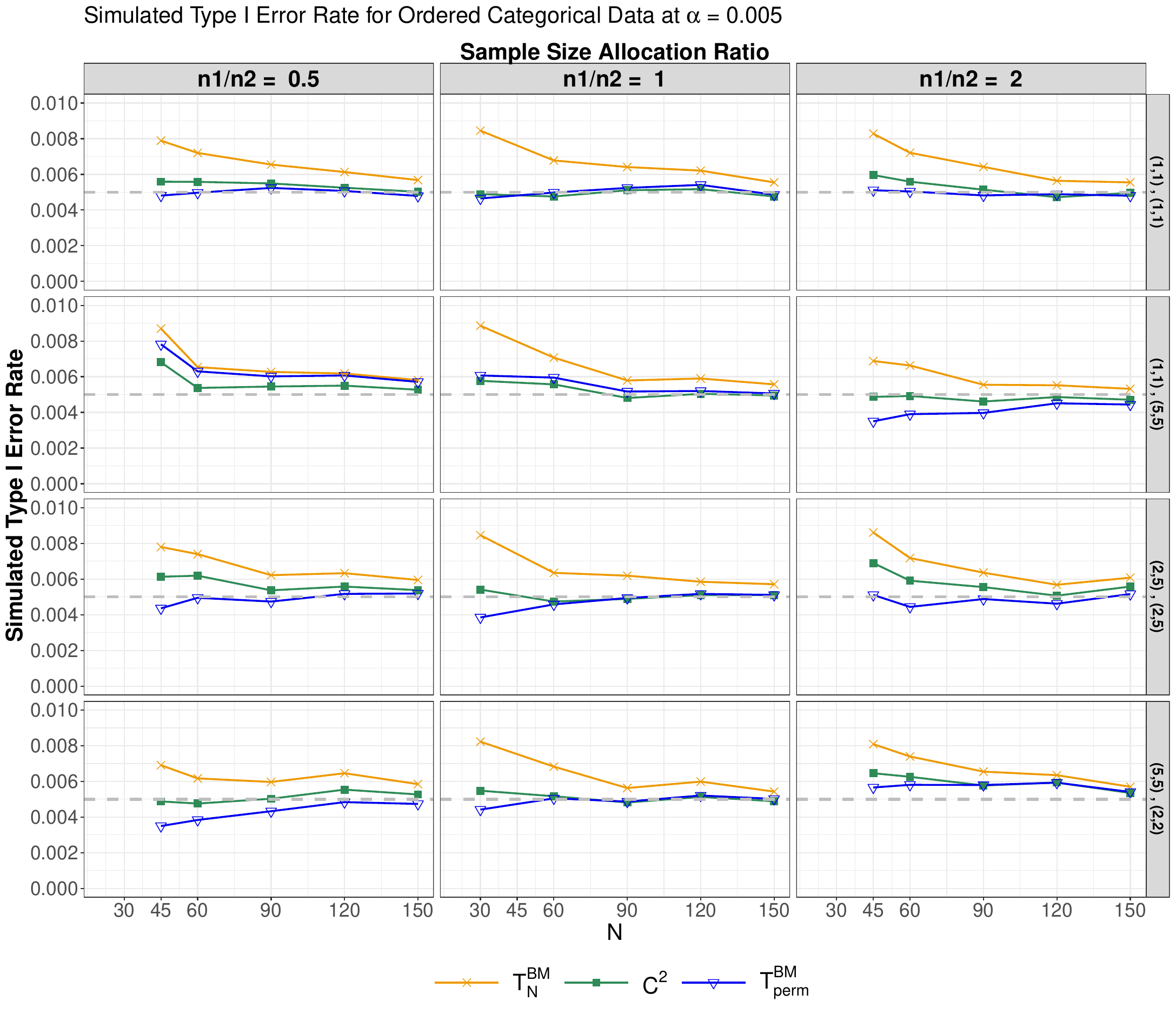}
    \caption{Type-I error rate for ordered categorical data settings 7 - 10 in Table~\ref{tab:settings-t1e} based on 100,000 replications at a two-sided nominal significance level of $\alpha = 0.005$ for all considered tests. The dashed grey line represents the nominal significance level.}
     \label{fig:ordinal_005}
\end{figure}
\newpage

\subsection{Power Simulation}
In the previous section, we examined the type-I error rate, illustrating that the proposed $C^2$-Test has decent small sample size properties. In practice, we are additionally concerned with the statistical power of the tests. Table~\ref{tab:settings-pwr} summarizes the settings we consider for power simulation. For the second group, we determine the distributional parameters $\mu, \alpha$ and $\lambda$ such that they correspond to some target Mann-Whitney effect $\tilde{\theta} > 0.5$. Regarding sample sizes, we consider $(n_1, n_2) \in \{(15, 15), (15, 30), (30, 15)\}$ to encompass both balanced and unbalanced designs. We again used 100,000 iterations in our simulation study. \newpage

\begin{table}[h]
    \centering
    \caption{Distributions considered in the power simulation study; All scenarios are chosen such that the alternative hypothesis $\mathcal{H}_1: \theta = \wttheta$ for some $\wttheta > 0.5$ holds}
    \begin{tabular}{p{1cm}p{4cm}p{4cm}p{4cm}} \toprule
        \textbf{Setting} & \textbf{Distribution} & \textbf{Control} & \textbf{Treatment} \\
        \midrule
1 & Normal distribution & $F_1 = \mathcal{N}(0,1)$ & $F_2 = \mathcal{N}(\mu,1)$ \\
2 & Normal distribution & $F_1 = \mathcal{N}(0,9)$ & $F_2 = \mathcal{N}(\mu,1)$ \\
\midrule
3 &  Ordered categorical data & Based on $F_1 = \mathcal{B}(1,1)$ & Based on $F_2 = \mathcal{B}(\alpha,1)$ \\
\midrule
4 & Exponential distribution & $F_1 = \mathcal{E}xp(\lambda)$ & $F_2 = \mathcal{E}xp(1)$ \\
\midrule
\bottomrule
\label{tab:settings-pwr}
    \end{tabular}
\end{table}

Since the results for the other distributions are virtually identical, we present only the power for settings 1 and 2 in the normal distribution case (see Table~\ref{tab:settings-pwr}) in the main manuscript. The results for the other settings are provided in the supplementary material. The $x$-axes represent the true Mann-Whitney effect $\theta$, while the $y$-axes display the simulated power. The grey dashed line indicates the conventionally chosen target power level of 80\%. Note that power curves appeared to be largely overlapping. \\

\begin{figure}[h]
    \centering
    \includegraphics[scale = 0.36]{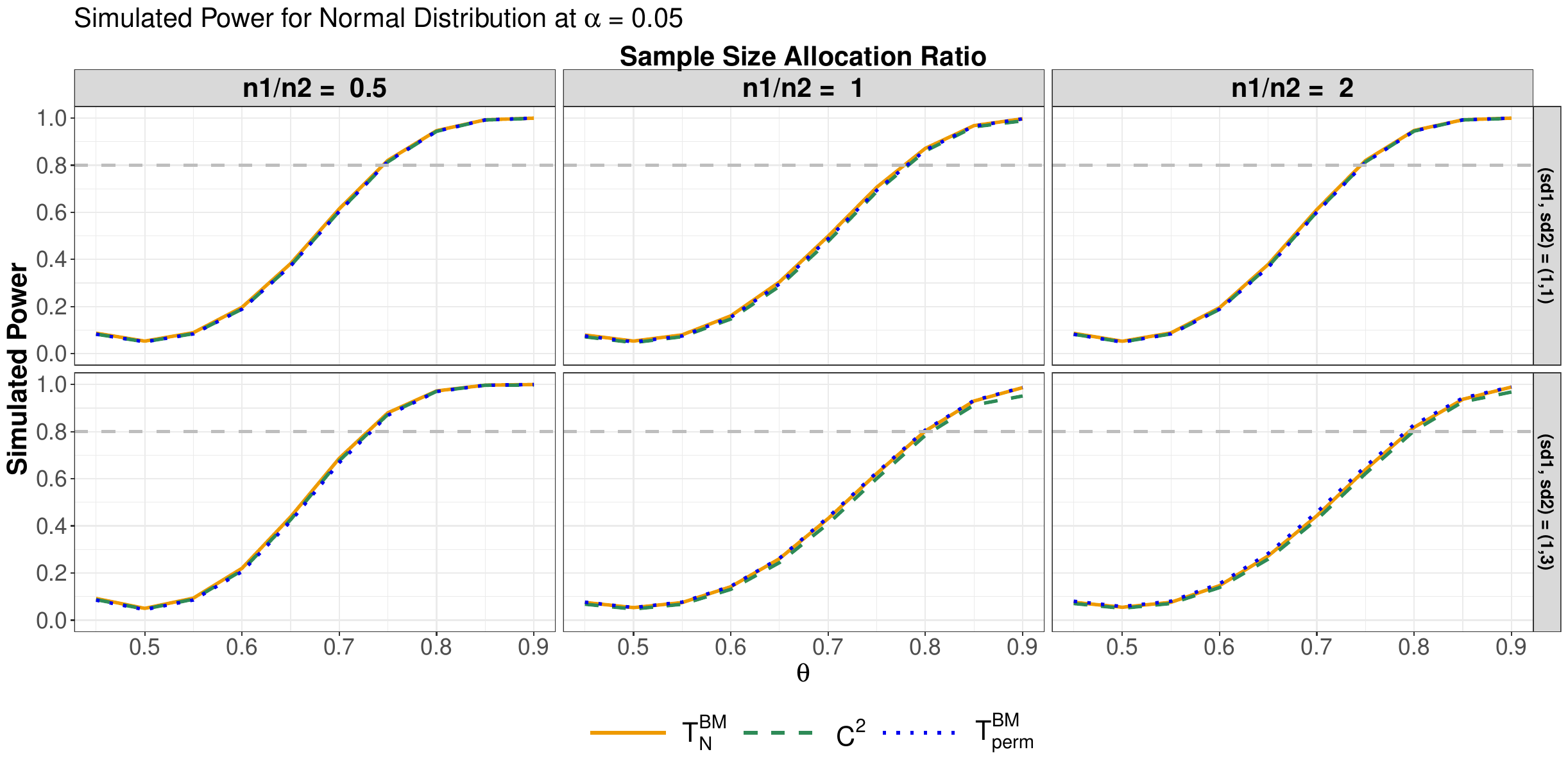}
    \caption{Power for normal distributions $F_1 = \mathcal{N}(0, \sigma_1^2)$ and $F_2 = \mathcal{N}(\mu, 1)$ based on 100,000 replications at a two-sided nominal significance level of $\alpha = 0.05$. $F_1$ and $F_2$ are chosen such that $\theta \in [0.45, 0.9]$.}
    \label{fig:pwr_normal}
\end{figure}

Figure~\ref{fig:pwr_normal} illustrates that differences in power among the tests are minimal, regardless of the variances or sample size configurations. Overall, the Brunner-Munzel and the permutation tests exhibit slightly higher power than the $C^2$-test, although these differences are negligible. Notable discrepancies occur only when $\sigma_1^2 < \sigma_2^2$ while $n_1 > n_2$, as highlighted in the bottom-right panel of Figure~\ref{fig:pwr_normal}. However, it should be noted that the slightly higher power of the Brunner-Munzel and permutation test likely stems from their slight liberality in scenarios involving negatively paired variances and small sample sizes, as illustrated in the bottom-right plots of Figure~\ref{fig:normal_005}. These findings conclude that the $C^2$-test achieves satisfactory power while providing improved control of the type-I error rate across the scenarios considered compared to the Brunner-Munzel test.

\subsection{Coverage Simulation}
\begin{figure}[htbp]
    \centering
    \includegraphics[scale = 0.35]{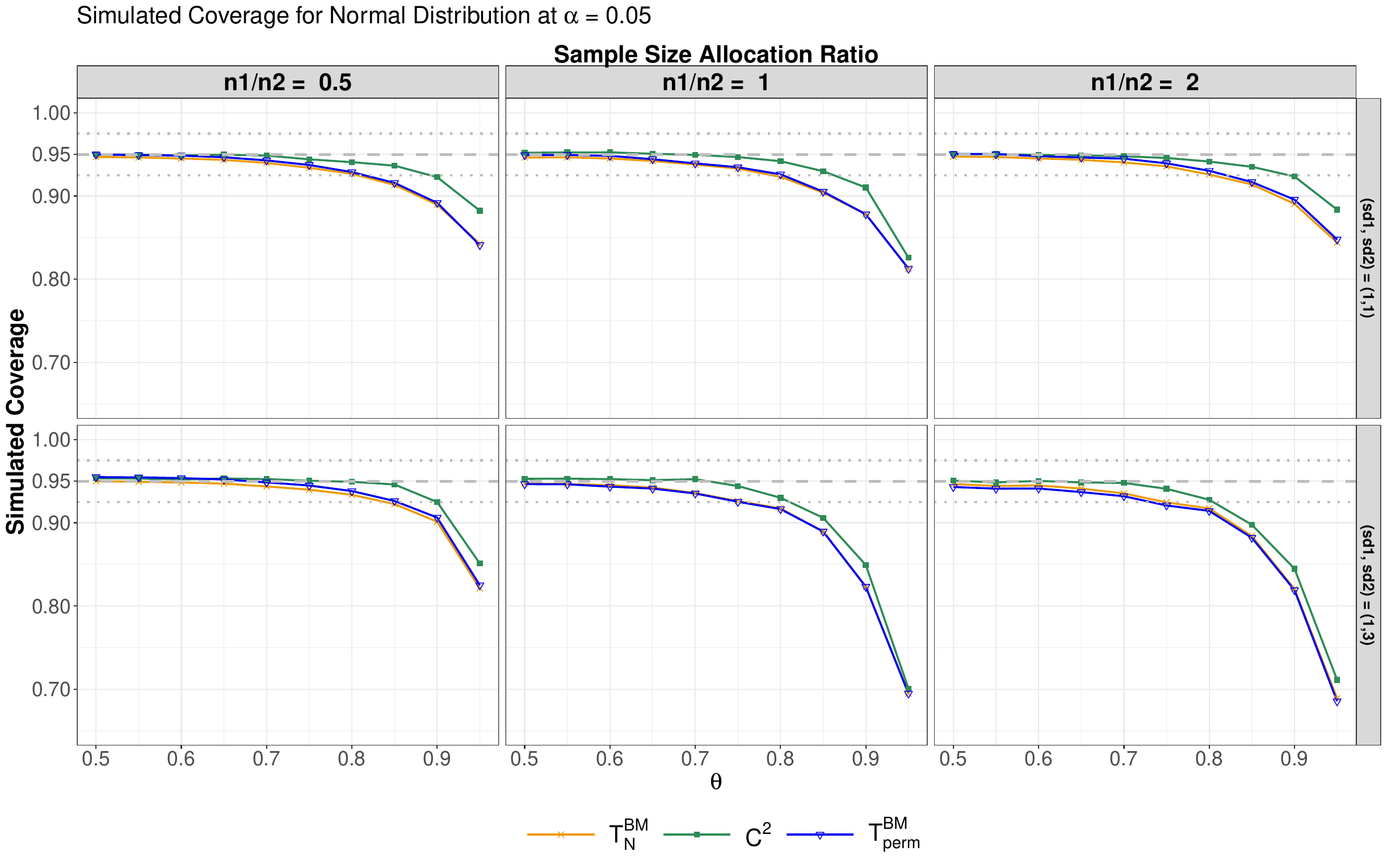}
    \caption{Simulated coverage probability for normal distributions $F_1 = \mathcal{N}(0, 1)$ and $F_2 = \mathcal{N}(0, \sigma_2^2)$ based on 100,000 replications at a two-sided nominal significance level of $\alpha = 0.05$ for all considered confidence intervals. Grey dotted lines represent the robust \citeintext{Bradley1978} limits $[0.925, 0.975]$.}
    \label{fig:coverage}
\end{figure}

Lastly, we briefly address the coverage probabilities of the confidence intervals in Equations~\eqref{eq:BM-CI}, \eqref{eq:Perm-CI}, and \eqref{eq:C2-CI}, derived from the Brunner-Munzel test, the studentized permutation test, and the $C^2$-test. To this end, we simulated normally distributed data as outlined in Table~\ref{tab:settings-pwr}, considering Mann-Whitney effects ranging from 0.5 to 0.95, ensuring consideration of rather extreme effects. The results are presented in Figure~\ref{fig:coverage} for $(n_1,n_2) = (15,30)$ (left plots), $n_1 = n_2 = 15$ (middle plots) and $(n_1, n_2) = (30,15)$ (right plots).

In Figure~\ref{fig:coverage}, all methods show a reasonable coverage at small values of $\theta$. However, as $\theta$ increases, all confidence intervals tend to become more liberal at some point. Specifically, the confidence intervals derived from inverting the Brunner-Munzel test and the studentized permutation test show a decline in coverage, falling below $0.925$ for approximately $\theta > 0.8$ in the homoscedastic case (upper row).  In contrast, the confidence interval associated with the $C^2$-test maintains higher coverage over a broader range of $\theta$. Notably, its coverage drops below $0.925$ only at $\theta > 0.85$ in the unbalanced case and $\theta > 0.9$ in the balanced case. A similar trend is observed in the heteroscedastic scenario (lower row), where the $C^2$-test also extends the range of $\theta$ for which coverage remains adequate, outperforming the alternative methods. In conclusion, our test provides a compatible confidence interval with improved coverage across a wider range of $\theta$ while additionally being range-preserving. Methods aimed at improving coverage specifically for extreme values of $\theta$ deserve particular attention, as emphasized by \citeintext{perme2019}. While the confidence interval based on the $C^2$-test already demonstrates an improvement over competing methods, addressing its behavior at the boundaries remains an open question for future research. \\

The results of the coverage simulations can also be interpreted in the context of diagnostic testing. To this end, note that the Mann-Whitney effect $\theta$ is known to be equivalent to the area under the ROC curve (AUC), a common measure of discriminatory accuracy in diagnostic testing (see, e.g., \citeauthor{bamber1975} (\citeyear{bamber1975}); \citeauthor{brunnerBook2019} (\citeyear{brunnerBook2019}), Section 2.2.3.3; \citeauthor{pauly2016} (\citeyear{pauly2016})). Although testing the null hypothesis of no discriminatory power, i.e. $\mathcal{H}_0: \theta = \nicefrac{1}{2}$, may often be of interest, \citeintext{pauly2016} point out that in diagnostic testing, it may also be of interest to evaluate hypotheses for any AUC value. Specifically, this approach is relevant when determining whether a new diagnostic test demonstrates improved accuracy over an existing one with an assumed $\theta_0 > 0.5$. To test the hypothesis $\mathcal{H}_0: \theta = \theta_0$, the test statistic 
\begin{align}
    C^2_{\theta_0} = \frac{\whtheta(1-\whtheta)}{\theta_0(1-\theta_0)\widehat{\sigma}_N^2}(\whtheta-\theta_0)^2
\end{align}
derived from Equation~\eqref{eq:der_C2} can be used and $\mathcal{H}_0: \theta = \theta_0$ is rejected if $C_{\theta_0}^2 > c_{1-\alpha}$. Since rejecting $\mathcal{H}_0: \theta = \theta_0$ is equivalent to $\theta_0 \notin [\theta_L^r, \theta_U^r]$, the simulation results for coverage (Figure~\ref{fig:coverage}) indicate that the test generally maintains the nominal $\alpha$ level for moderately large values of $\theta_0$.


\section{Analysis of the Example}
\label{sec:application}
In this section, we revisit the data from the shoulder tip pain study, which aimed to determine whether a specific suction procedure reduces pain in patients after laparoscopic surgery more effectively than the standard of care. In this context, values of $\theta > 0.5$ indicate a tendency towards lower pain scores in the treatment group, suggesting the procedure's effectiveness, while values of $\theta < 0.5$ favor the control group. 
\begin{table}[h]
\renewcommand{\arraystretch}{1.4} 
\centering
    \caption{Results of the pain score study}
    \begin{tabular}{|l|c|c|c|} 
        \hline
         & \multicolumn{3}{c|}{Test Procedure} \\[0.15cm]
         \cline{2-4}
         Parameter &  Brunner-Munzel Test & Permutation Test  & $C^2$-Test \\
         \hline 
         $\whtheta_N$ & 0.837 & 0.837 & 0.837 \\
         Variance estimate & $\widehat{\nu}_{DL}^2 = 0.172$ & $\widehat{\nu}_{DL}^2 = 0.172$  & $N\widehat{\sigma}_{N}^2 = 0.171$ \\
         95\%-Confidence interval & $[0.704, 0.970]$ & $[0.708, 0.970]$ & $[0.677, 0.927]$ \\
         Test statistic & 5.20 & NA  & 14.9\\
         $p$-Value  & 0.0000187 & 0 & 0.000116 \\
         \hline
         Decision & reject $\mathcal{H}_0$ & reject $\mathcal{H}_0$  & reject $\mathcal{H}_0$ \\
         \hline
    \end{tabular}
\end{table}
Using this data, we obtain a Mann-Whitney effect estimate of $\whtheta_N = 0.837$, indicating that pain scores tend to be lower in participants undergoing the specific junction procedure. The estimated probability for ties in this dataset is $\widehat{\tau}_N= 0.182$. We observe that the variance estimate for the Brunner-Munzel test ($\widehat{\nu}_{DL}^2 = 0.172$) and the unbiased variance estimate used in the $C^2$-test ($\widehat{\sigma}_N^2 = 0.171$) take comparable values. Regarding the confidence intervals, we observe a slightly larger upper bound for the interval corresponding to the Brunner-Munzel test ($[0.704, 0.970]$) compared to a slightly smaller lower bound for the interval corresponding to the $C^2$-test ($[0.677, 0.927]$). Despite these minor differences, all three tests consistently lead to the conclusion that the null hypothesis $\mathcal{H}_0: \theta = \nicefrac{1}{2}$ can be rejected at a nominal significance level of $\alpha = 0.05$ in favor of the treatment group, indicating that the specific suction procedure significantly reduced the pain scores in comparison to the control treatment. 

\section{Discussion and Outlook}
\label{sec:discussion}
In our paper, we proposed a novel test for the nonparametric Behrens-Fisher problem, which approximates the true variance $\sigma_N^2$ by employing the ratio of an unbiased estimator to the theoretical upper bound of the true variance, resulting in an adjusted version of the Brunner-Munzel test. Furthermore, we provided compatible range-preserving confidence intervals for the new test. \\[-0.1cm]

In our simulation study, we observed that the $C^2$-test demonstrates improved type-I error rate control in small sample sizes compared to the Brunner-Munzel test across a broad range of data-generating mechanisms, even at significance levels as stringent as $\alpha = 0.005$. Hence, the presented methods align with recent recommendations questioning the use of $\alpha = 0.05$ for hypothesis testing, indicating that the derived test should be favored over the Brunner-Munzel test. Furthermore, the proposed test was comparable to the studentized permutation test by \citeintext{neubert2007} regarding type-I error control. While the studentized permutation test slightly exceeded the nominal $\alpha$-level in case of negatively paired variances, the proposed tests struggled slightly under highly skewed ordered categorical data conditions. In terms of power, the differences were mainly negligible. Regarding coverage of the confidence intervals, we observed that the proposed confidence interval maintains adequate coverage over a broader range of $\theta$ compared to the competing methods. Equivalently, one can conclude that the test for the hypothesis $\mathcal{H}_0: \theta = \theta_0$ appropriately controls the nominal $\alpha$-level within this range, qualifying it for use in diagnostic testing. Finally, the proposed method can be expressed in closed-form mathematical formulas rather than relying on resampling, making it more efficient in terms of runtime, which may be a favorable feature when conducting simulation-based study planning. With the favorable simulation results, we conclude that our test should be considered for use in future work. \\[-0.1cm]

While we've outlined its desirable properties, our method is not without limitations. First, it relies on the asymptotic normality of $\whtheta$, which may be questionable for small sample sizes such as $n < 15$ per group. In those cases, we recommend computing p-values and confidence intervals using studentized permutation, which is robust even for sample sizes $n<15$ \citebrackets{pauly2016}. Second, similarly to the competitor methods, when effect sizes approach boundary values, the proposed confidence intervals exhibit liberal coverage probabilities. Therefore, confidence intervals for small sample sizes and large effects should be interpreted cautiously. Addressing this issue in the presence of extreme effects remains open question for future research. \\[-0.1cm] 

Alongside investigating properties under extreme effects, our work lays the foundation for numerous further research directions. First, our method distinguishes itself from established approaches by incorporating the unbiased variance estimator proposed by \citeintext{brunner2024_2}. When combined with range-preserving transformations, such as logit or probit transformations, this estimator could be integrated into modified studentized permutation procedures to potentially enhance the performance of the tests and confidence intervals proposed in \citeintext{pauly2016}, particularly for very small sample sizes. However, this paper intends to focus on tests that can be expressed through explicit formulas. Additionally, incorporating the Birnbaum-Klose inequality to improve adherence to type-I error rate control could be leveraged to refine methods in other fields of application. For example, in diagnostic testing, a common research question is to compare biomarkers with respect their ability to distinguish between diseased and non-diseased individuals, i.e. to test $\mathcal{H}_0: \theta_1 = \theta_2$, where $\theta_1$ and $\theta_2$ represent Mann-Whitney effects referring to the two biomarkers. A straightforward adaptation could replace the variance estimator in the commonly used DeLong test with that of \citeintext{brunner2024_2}. In contrast, techniques similar to those used for the $C^2$-test could be developed to investigate alternative methods, particularly for small sample sizes. Another promising research area involves generalized pairwise comparisons (GPC), as introduced by \citeintext{buyse2010}, which extend standard nonparametric methods to multivariate prioritized outcomes using $U$-statistics theory. Adapting our test proposals to a GPC framework should be relatively straightforward, as estimators such as $\whtheta_N$ and $\widehat{\sigma}_N^2$ should remain valid when derived from $U$-statistics with count functions rather than empirical distribution functions and ranks. However, establishing the theoretical underpinnings of this approach remains an open challenge. Finally, practical considerations such as sample size planning for the $C^2$-test or the extension to several samples and paired data represent important areas of study.

\newpage

\subsection*{Data availability statement}
Original data presented in Section~\ref{sec:example} was analyzed. The data was taken from \citeintext{lumley1996} and also analyzed in the papers of \citeintext{brunner2000} and \citeintext{neubert2007}. All code used to reproduce the results in this paper will be uploaded upon revision for review by the Responsible Research Editor and will be made publicly available on GitHub. We intend to integrate the methods into the R package rankFD upon publication.

\subsection*{Conflict of interest}

The authors declare no potential conflicts of interest with respect to the research, authorship and/or publication of this article.

\subsection*{Funding}
This work was supported by the German Research Foundation (grant KO 4680/4-2).

\subsection*{Authors' contributions}
StS contributed to developing the methods, wrote the manuscript, coded, and created the tables and figures. FK provided critical review of the manuscript and supervised the simulation study. EB  proposed the paper project, suggested to evaluate the ratio method, contributed to the development of the methods and provided supervision of the project. All authors read and approved the final manuscript.

\subsection*{Acknowledgements}
The authors would like to thank the German Research Foundation for supporting this research by the grant KO 4680/4-2.

\newpage

\section*{Appendix}
\subsection*{A.1 Derivation of the Ratio Method Confidence Interval}
We derive the ratio method confidence intervals $\theta_{L,U}^r$ in Equation~\eqref{eq:C2-CI} by solving the the quadratic inequality
\begin{align*}
    (\whtheta_N-\theta)^2 \geq \whq c_{1-\alpha}\theta(1-\theta),
\end{align*}
where $\whq = \frac{\widehat{\sigma}_N^2}{\whtheta_N(1-\whtheta_N)}$,
for $\theta$:
\begin{align*}
   &\hspace{1.45cm} (\whtheta_N-\theta)^2 \geq \whq c_{1-\alpha}\theta(1-\theta) \\
   &\Longleftrightarrow  \qquad   \underbrace{(1 + \whq c_{1-\alpha})}_{=a}\theta^2 \underbrace{- (2\whtheta_N + \whq c_{1-\alpha})}_{=b}\theta + \underbrace{\whtheta_N^2}_{=c} \geq 0 \\
   &\Longleftrightarrow  \qquad a\theta^2 + b\theta + c \geq 0.
\end{align*}
Solving this quadratic inequality results in the solutions
\begin{align*}
    \theta_{L,U}^r &=  \frac{\left(2\whtheta_N + \whq c_{1-\alpha}\right) \mp \sqrt{\left(2\whtheta_N + \whq c_{1-\alpha}\right)^2 - 4\left(1 + \whq c_{1-\alpha}\right)\whtheta_N^2}}{2\left(1 + \whq c_{1-\alpha}\right)} 
\end{align*}
By further simplification, we obtain that
\begin{align*}
      \theta_{L,U}^r
        &= \frac{1}{2\left(1 + \whq c_{1-\alpha}\right)} \left(2\whtheta_N + \whq c_{1-\alpha} \mp \sqrt{\whq^2 c_{1-\alpha}^2 + 4\widehat{\sigma}_N^2c_{1-\alpha}}\right)
\end{align*}
which are the boundaries given in Equation~\eqref{eq:C2-CI}.
\newpage

\subsection*{A.2 Proof that the Confidence Interval $\theta_{L,U}^r$ in \eqref{eq:C2-CI} is range-preserving}
In Remark 4.2, we stated that the confidence interval $\theta_{L,U}^r$ is range-preserving. We will prove this property by contradiction. First, recall that the confidence interval is given by
\begin{align*}
        \theta_{L,U}^{r}
        &= \frac{1}{2\left(1 + \whq c_{1-\alpha}\right)} \left(2\whtheta_N + \whq c_{1-\alpha} \mp \sqrt{\whq^2 c_{1-\alpha}^2 + 4\widehat{\sigma}_N^2 c_{1-\alpha}}\right),
\end{align*}
where $\whq = \frac{\widehat{\sigma}_N^2}{\whtheta_N(1-\whtheta_N)}$. If the confidence interval were not range preserving, it must hold that $\theta_U^r > 1$, that is,
\begin{align*}
    2\whtheta_N + \sqrt{\whq^2 c_{1-\alpha}^2 + 4\widehat{\sigma}_N^2 c_{1-\alpha}} &> 2 + \whq c_{1-\alpha}
\end{align*}
which can be rewritten to
\begin{align*}
    c_{1-\alpha} \widehat{\sigma}_N^2 > (1 - \whtheta_N)^2 + (1 - \whtheta_N) \whq c_{1-\alpha}
\end{align*}
by straightforward computations. Since $c_{1-\alpha} \widehat{\sigma}_N^2 = \whq c_{1-\alpha} \whtheta_N (1 - \whtheta_N) $, and excluding the case $\whtheta_N = 1$, it follows that
\begin{align*}
    \whq c_{1-\alpha} \whtheta_N > (1 - \whtheta_N) + \whq c_{1-\alpha},
\end{align*}
which simplifies to
\begin{align*}
    0 > (1 - \whtheta_N) + \whq c_{1-\alpha} (1 - \whtheta_N),
\end{align*}
which is a contradiction, as both $\whq, c_{1-\alpha}$ are positive, and (by assumption) $\whtheta_N \in [0, 1)$. By similar arguments (and excluding $\whtheta_N = 0$) it follows that $\theta_L^r > 0$. For $\widehat{\theta}_N \in \{0,1\}$, we use the confidence interval in Equation~\eqref{eq:CI_theta=1} and \eqref{eq:CI_theta=0}, which are range-preserving by construction.
\newpage

\subsection*{A.3 Simulation of Ordered Categorical Data}
To simulate ordinal data, we partition the data into $J$ categories $\mathcal{C}_1 < \ldots < \mathcal{C}_J$, where a higher index signifies a more favorable outcome. Following the approach outlined by \citeintext{brunner2021} and \citeintext{nowak2022}, we initiate the simulation by generating data from a Beta-distribution. Let $Y_{ik} \sim \mathcal{B}(\alpha_i, \beta_i)$, where $i \in \{1,2\}$ and $k = 1,\dots,n_i$, denotes a random variable following a Beta-distribution with parameters $\alpha_i > 0$ and $\beta_i > 0$. These parameters are chosen such that:
\begin{align*}
\mathbb{E}[Y_{ik}] &= \frac{\alpha_i}{\alpha_i + \beta_i} \\
\mathbb{V}[Y_{ik}] &= \frac{\alpha_i \beta_i}{(\alpha_i + \beta_i)^2(\alpha_i + \beta_i + 1)}.
\end{align*}
Using this, we create ordered categorical data $X_{ik}, i \in \{1,2\}, i = 1,\dots,n_i$, according to the intervals:
\begin{align*}
X_{ik} = \mathcal{C}_j \quad \text{if} \quad Y_{ik} \in \left[\frac{(j-1)}{J}, \frac{j}{J}\right)
\end{align*}
for $j=1,\dots,J$. This results in the probability mass function:
\begin{align*}
\text{P}(X_{ik} = \mathcal{C}_j) = \text{P}\left(Y_{ik} \in \left[\frac{(j-1)}{J}, \frac{j}{J}\right)\right).
\end{align*}
To mimic the commonly employed 5-point Likert scale data, we set $J = 5$ such that
\begin{align*}   
X_{ik} = 
     \begin{cases}
         \mathcal{C}_1 &\quad \text{if} \quad Y_{ik} \in [0, 0.2)\\
         \mathcal{C}_2 &\quad \text{if} \quad Y_{ik} \in [0.2, 0.4)\\
         \mathcal{C}_3 &\quad \text{if} \quad Y_{ik} \in [0.4, 0.6)\\
         \mathcal{C}_4 &\quad \text{if} \quad Y_{ik} \in [0.6, 0.8)\\
         \mathcal{C}_5 &\quad \text{if} \quad Y_{ik} \in [0.8, 1].
     \end{cases}
\end{align*}
By adjusting $\alpha_i$ and $\beta_i$, it is possible to construct both homoscedastic and heteroscedastic data distribution settings. Furthermore, we can select $\alpha_i$ and $\beta_i$ to match a target Mann-Whitney effect $\tilde{\theta}$, i.e. to fulfill $\int F_1 dF_2 = \tilde{\theta}$. In our power simulations, we achieve this by fixing three of the four parameters and solving for the remaining parameter to attain the target effect. The following relation facilitates this. With $J$ denoting the number of categories and $0=c_0 < c_1 < \dots < c_{J-1} < c_J = 1$ the cutoff-values, the Mann-Whitney effect is given through
\begin{align*}
    \theta &= \text{P}(X_{2i} > X_{1j}) + \nicefrac{1}{2}\text{P}(X_{2i} = X_{1j}) \\
      &= \sum_{i=2}^{J}\left((F_{\alpha_2, \beta_2}(c_i) - F_{\alpha_2, \beta_2}(c_{i-1}))\left(\sum_{l=1}^{i-1}(F_{\alpha_1, \beta_1}(c_l) - F_{\alpha_1, \beta_1}(c_{l-1}))\right)\right) \\ &+ 
      0.5 \sum_{i=1}^{J}(F_{\alpha_1, \beta_1}(c_i) - F_{\alpha_1, \beta_1}(c_{i-1}))(F_{\alpha_2, \beta_2}(c_i) - F_{\alpha_2, \beta_2}(c_{i-1}))
\end{align*}
\newpage  

\subsection*{A.4 Seperate Sample Handling}
In case of completely separated samples $\mathbf{X}_1 = (X_{11},\dots,X_{1n_1})$ and $\mathbf{X}_2 = (X_{21},\dots,X_{2n_2})$ (e.g. if $\min\{\mathbf{X}_1\} > \max\{\mathbf{X_2}\}$ or vise versa), we have that $\whtheta_N \in \{0,1\}$ and $\widehat{\sigma}_N^2 = 0$ or $\widehat{\nu}_{DL}^2 = 0$, leading to undefined values of all test statistics under consideration. We must treat this case separately to address this singularity in our simulation study. Here, we briefly introduce our exception-handling approach for all methods.

\subsubsection*{A.4.1 $C^2$-Test}
For the $C^2$-test, the case of separate samples has already been addressed in the main manuscript, where we proposed replacing the ratio approximation with the maximum variance estimate, $\sigma_{max}^2 = \theta(1-\theta)/m$, thereby eliminating the need for variance estimation.
\subsubsection*{A.4.2 Brunner-Munzel Test and Studentized Permutation Test}
In the case of separate samples, the variance estimate $\widehat{\nu}_{DL}^2$ becomes zero, leading to an undefined value of $T_N^{BM}$ as well as undefined confidence bounds for both the Brunner-Munzel test and the studentized permutation test. Here, we employed the ``one-step-back'' method as described in \citeintext{brunnerBook2019} (Section 3.5.3).  This method adjusts $\whtheta_N$ for extreme cases ($\whtheta_N = 0$ or $\whtheta_N = 1$) by selecting the closest possible values to $0$ or $1$ and the smallest non-zero variance estimator, based on minimal overlap (nearly separated samples). Concretely:  
\begin{itemize}
        \item When no ties occur, $\whtheta_N = 1 - 1/(n_1n_2)$ is the closest value to $1$, and $\whtheta_N = 1/(n_1n_2)$ is the closest value to $0$. The smallest non-zero variance estimator is given by $\widehat{\nu}_{DL}^2 = 2N/(n_1^2n_2^2)$.  
        \item In the presence of ties, the smallest overlap corresponds to a single tied value. In this case, $\whtheta_N = 1 - 1/(2n_1n_2)$ is closest to $1$, and $\whtheta_N = 1/(2n_1n_2)$ is closest to $0$. The smallest non-zero variance estimator is then $\widehat{\nu}_{DL}^2 = N/(2n_1^2n_2^2)$.  
\end{itemize}

\bibliography{main}
\newpage

\end{document}